\documentclass[twocolumn,showpacs,amsmath,amssymb]{revtex4}

\usepackage{graphicx}
\usepackage{color}

\begin{document}

\title{Acoustic Emission from Breaking a Bamboo Chopstick}
\author{Sun-Ting Tsai, Li-Min Wang, Panpan Huang$^{\dagger}$, Zhengning Yang$^{\ddagger}$, Chin-De Chang, and Tzay-Ming Hong$^{\ast}$}
\affiliation{Department of Physics, National Tsing Hua University, Hsinchu 30013, Taiwan, Republic of China}
\date{\today}
\begin{abstract}

The acoustic emission from breaking a bamboo chopstick or a bundle of spaghetti is found to exhibit similar behavior as the famous seismic laws of Gutenberg-Richter,  Omori, and B{\aa}th. By use of a force-sensing detector, we establish a positive correlation between the statistics of sound intensity and the magnitude of tremor. We also manage to derive these laws analytically without invoking the concept of phase transition, self-organized criticality, or fractal. Our model is deterministic and relies on the existence of a structured cross section, either fibrous or layered. This success at explaining the power-law behavior supports the proposal that geometry is sometimes more important than mechanics.

\end{abstract}
\maketitle

Fracture\cite{fracture} is a complex phenomenon with many interesting properties, such as crack propagation\cite{crack}, breakdown\cite{stanley}, and self-affine fractals in the crack surface morphology\cite{selfaffine}. This kind of studies started in seismology and attracted the attention of physicists and material scientists. Crackling noise\cite{sethna,salje} in the acoustic emission from fracturing wood plate\cite{plate}, paper\cite{paper,santucci}, rock\cite{rock}, concrete\cite{concrete}, and charcoal\cite{charcoal} has been found to exhibit similar statistical properties as earthquake\cite{rupture,fractal}.
On the other hand, the slow growth of a single crack in a fibrous sheet has been studied by the lattice model\cite{santucci} which incorporates thermodynamics to describe the temperature dependence.  To understand the failure process, the fiber bundle model\cite{bundle} has also been extended\cite{kun} by introducing time-dependent damage accumulation of fibers to capture the stochastic nature of fracture. 

In contrast to 2- and 3-D samples in previous fracture experiments, we shall adopt the 1-D like chopstick because it rarely fractures in multiple places at the same time, which makes the interpretation of data easier. 
Chopsticks are easily accessible because they are indispensable implements for most Asian families and carry special cultural meanings\cite{chop} in Chinese tradition. They are usually made by bamboo in Taiwan and China. Bamboo exhibits a fibrous cross section. Its excellent compressive strength, endowed from the vascular bundles scattered throughout the stem in the cross section instead of in a cylindrical arrangement, enables us to collect about 400 crackling sounds from breaking one chopstick, as opposed to just one sound from a chalk or twig.

\begin{figure}[!h]
  \centering
  \includegraphics[width=8cm]{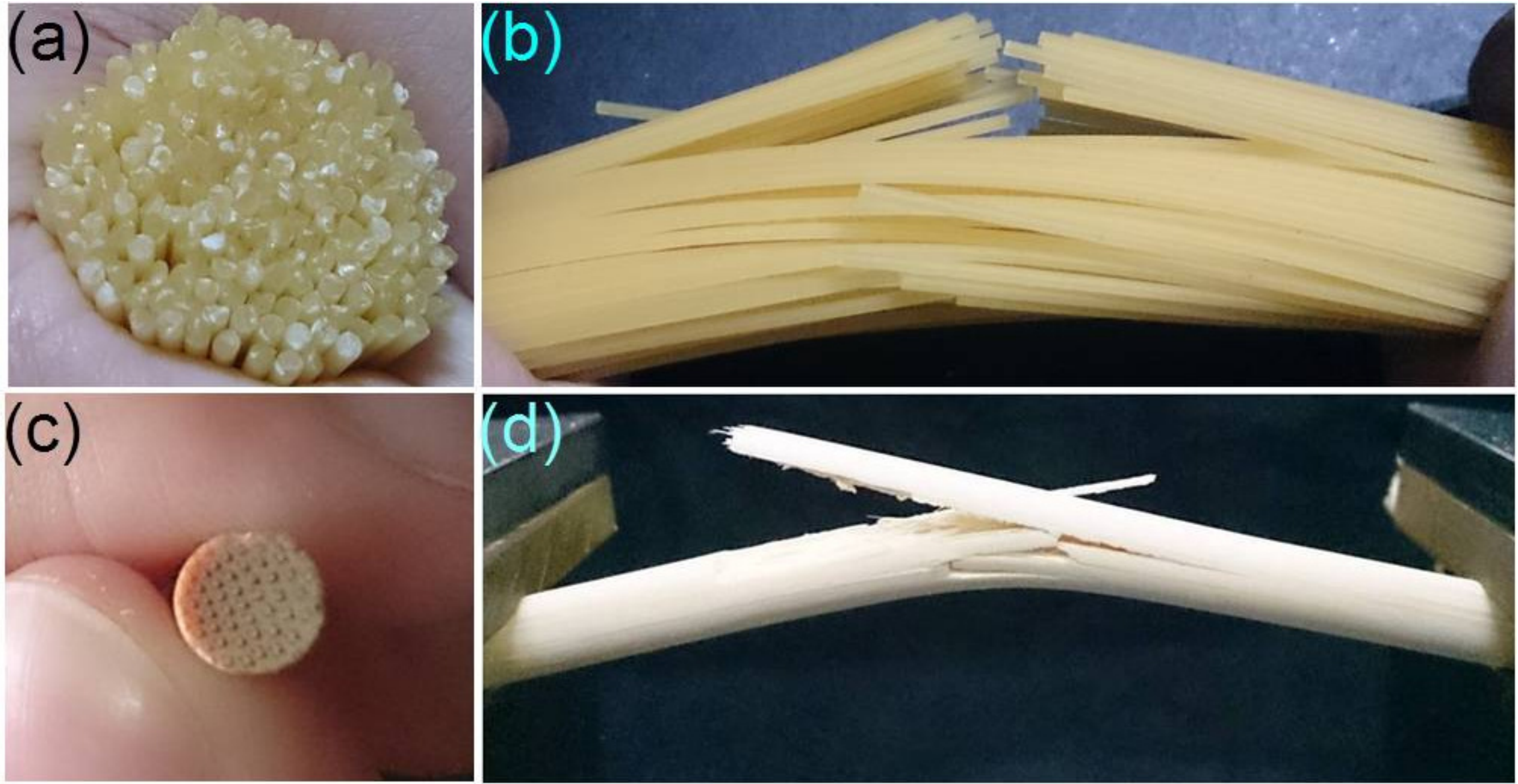} 
  \caption
  {(color online) Panels (a) and (c) show the similar structure in cross section between a bundle of spaghetti and a bamboo chopstick. Panels (b) and (d) are the side view during fracture.
}\label{spaghetti} 
\end{figure}

\begin{figure}[!h]
  \centering
  \includegraphics[width=8cm]{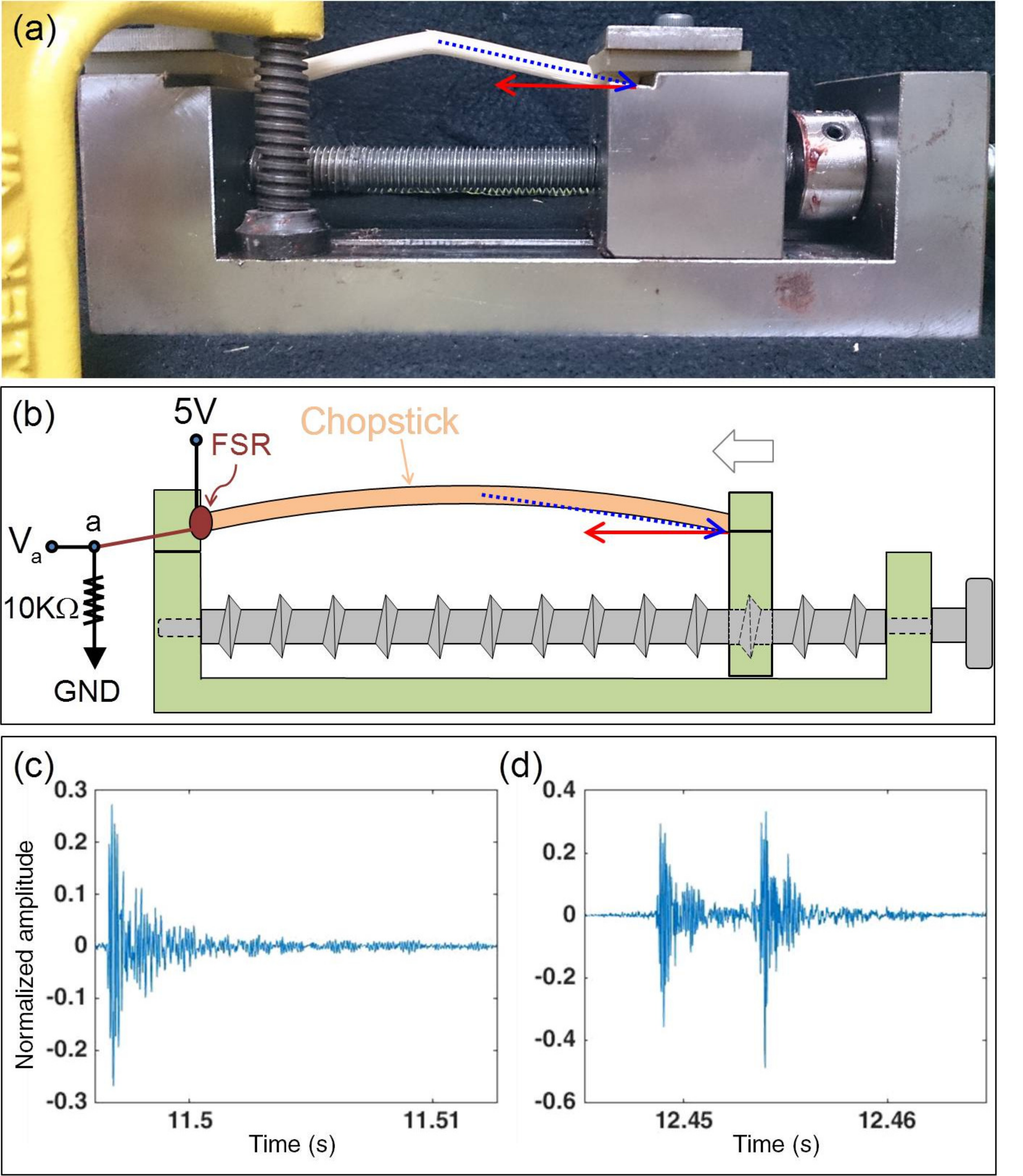} 
  \caption
  {(color online) Panels (a,b) show the fracture machine. FSR is the acronym for force-sensing resistor, and GND is the shorthand for ground. The blue dotted arrow  represents the arm of force, and the red solid arrow the force exerted on one end of the chopstick. Panel (c) shows a typical sound pulse, and (d) an example when one long pulse or two overlapping pulses need to be distinguished.
}\label{setup} 
\end{figure}

Our main sample is the cheap and omnipresent bamboo chopstick of diameter 0.67cm and length 20.5cm. Sticks are also taken directly from the bamboo tree to make sure our conclusions are independent of the length, freshness, and processing to turn them into chopsticks. A bundle of spaghetti\cite{youtub} is also used to check the generality of our model because it shares similar cross section as bamboo (see Fig.\ref{spaghetti}). The fracture machine in Fig.\ref{setup} is placed inside a soundproof chamber with foam rubber plank on the interior to avoid echo. A Sony ECM166BC microphone that is connected to a Sony ICD-PX333 recorder picks up fracture sound. By tightening the screw, the movable metal block  was driven and compressed the chopstick. The bolted plate is to secure the bent chopstick from slipping and flying off. The chopstick was flipped into a horizontal position as soon as it became bent to prevent fracturing near the plate edge. Crackling sound was recorded at a sample rate of 44100 points per second in 16-bit precision. The amplitude was measured in computer units and the maximum amplitude $A_{\rm max} =2^{15}-1$. The gain of sound card was constant and samples were positioned at a distance of 5 cm from the microphone. We kept the process of fracture at a steady velocity of about 200 s per chopstick.

Average amplitude of background noise was $3\times10^{-3}$ as normalized by $A_{\rm max}$, and thrice this amount was set as the noise threshold. The c code algorithm automatically integrated the sound intensity every 200/44100
s. When the value exceeded background noise, beginning of a new pulse was marked,  as in Fig.\ref{setup}(c). Whenever a dilemma arose at distinguishing a long pulse from two overlapping pulses as Fig.\ref{setup}(d), we resorted to a smaller time step to examine the intermediate area by including just six amplitude peaks of the oscillation within the pulse(s). Since this value is expected to decrease as a pulse fades, a sudden switch to increasing function indicates the beginning of a second pulse.

\begin{figure}[!h]
  \centering
  \includegraphics[width=8cm]{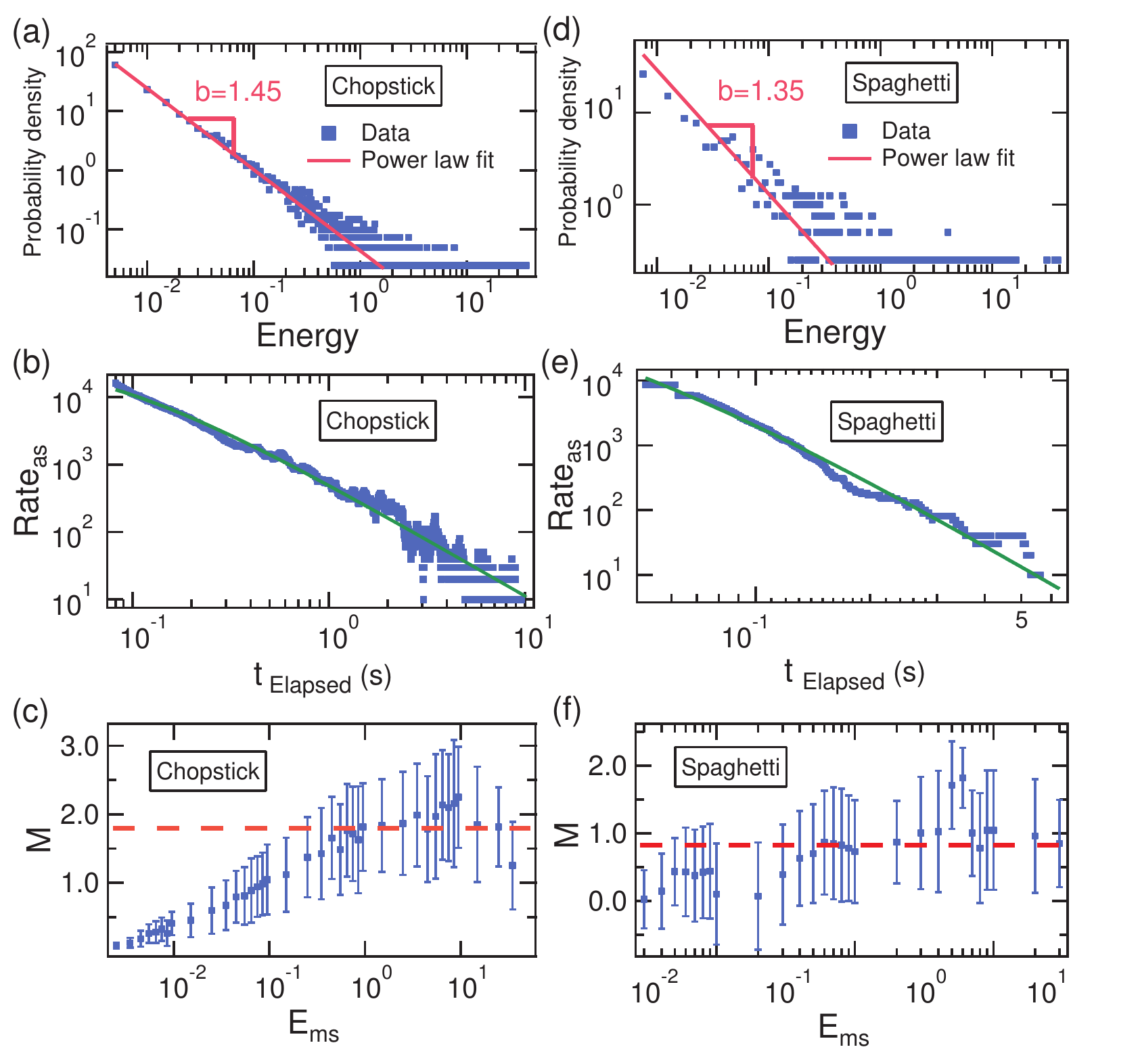} 
  \caption
  {(color online) Panels (a, b, c) are for crackling sound from the bamboo chopstick, while  (d, e, f) from a bundle of spaghetti. They mimic the Gutenberg-Richter law, Omori's law, and B{\aa}th's law for earthquake. The power-law exponent in (a) and (d) is determined to be 1.45 and 1.35.  The exponent and shift in (b) and (e) are ($p=1.68$, $c=0.07$) and (3.53, 0.027). The relative magnitude M in panels (c) and (f) is defined as the average of $\log(E_{\rm ms})-\log(E_{\rm la})$ where $E_{\rm ms}$ and $E_{\rm la}$ denote respectively the energy of main shock and its largest aftershock. Error bars represent standard deviations.
}\label{data} 
\end{figure}

After squaring the amplitude and integrating over the duration time to obtain the energy for each sound pulse, we grouped the data into a histogram of probability function versus pulse energy. A full-logarithm was then taken in Fig.\ref{data}(a) and (d) to reveal a good alignment, indicative of a power law: $P(E)\sim 1/E^b$ that is reminiscent of the Gutenberg-Richter law\cite{gutenberg}. The exponent, $b$, for chopstick is  about 1.45  by use of the maximum likelihood estimation (MLE). When we sliced the chopstick horizontally into two halves, the power law persisted but the upper (lower) hemisphere became  $b=1.48$ (1.40). The empirical value of $b$ for earthquake is commonly close to 1.0 in seismically active regions, but may vary in the range of 0.5 and 2 depending on the source environment of the region\cite{schorlemmer}.

The aftershock rate in Fig.\ref{data}(b) and (e) was defined as the number of aftershocks divided by the time window of 100ms, against the elapsed time, $t_{\rm Elapsed}$. And aftershocks referred to the sound pulses on the time series between two local maxima in pulse energy; i.e., the main shocks. Since the number of local maxima roughly matches that of vascular bundles, we believe the fracture of each bundle gives rise to one main shock and a few aftershocks.  A shifted power law, rate$=k/(c+t_{\rm Elapsed})^p$ where $k$ and $c$ are constants, similar to the Omori's law\cite{omori} was deduced. The exponent in Fig.\ref{data}(b) was determined by MLE to be $p=1.68$, slightly larger than the region of 0.75-1.5 set for earthquakes. Finally, after an initial dip in Fig.\ref{data}(c) and (f) that is known to exist in real data\cite{bath3,bath2}, the M-value became independent of the main-shock magnitude and saturated at 1.7-1.8 in Fig\ref{data}(c) - somewhat higher than 1.1-1.2 for the B{\aa}th's law\cite{bath}.

In spite of a multitude of workers who have compared the acoustic emission from laboratory material to the shock energy of earthquakes,  the reasonable assumption that these two energies are correlated was rarely proven. We thus decided to affix a force-sensing resistor (FSR) to the end of chopstick in Fig.\ref{setup} and compared the timing and statistical behavior of the discrete events of tremor and acoustic emission.  When the reaction force by the chopstick increased, the resistance of FSR (Interlink Electronics FSR402) decreased from an initial value higher than 10M ohm and generated a nonzero voltage, $V_{a}$. An Arduino UNO board then collected the measurement of $V_{a}$ at a sample rate of 100 points per second.

\begin{figure}[!h]
  \centering
  \includegraphics[width=8cm]{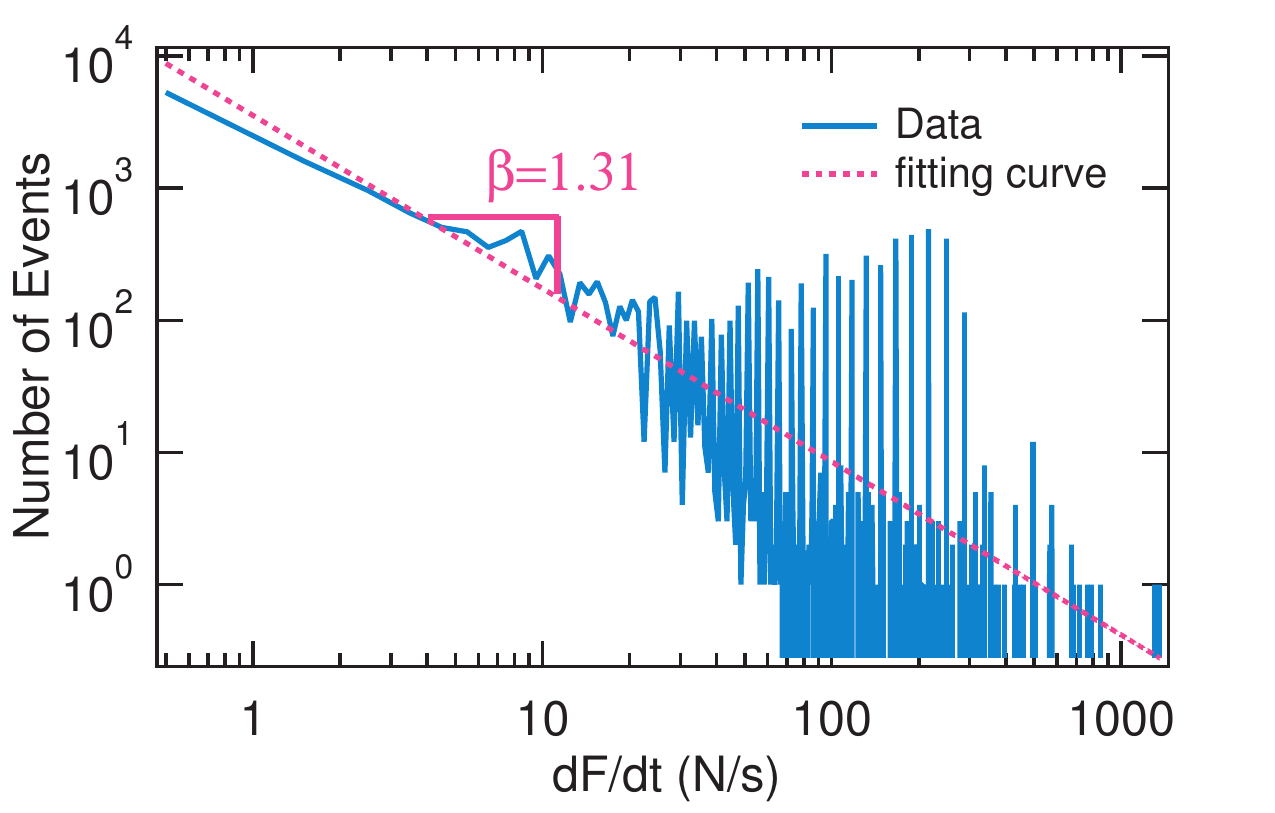} 
  \caption
  {(color online) Full-log plot of the number of events versus the magnitude of force change picked up by FSR. Data were fit to a power law (the red dotted line) with an exponent roughly equal to that of crackling sound in Fig.\ref{data}(a).
}\label{FSR} 
\end{figure}

Any fracture is expected to cause a  sudden drop in reaction force against the FSR, but not necessarily accompanied by a detectable sound. Therefore, there is no one-to-one correspondence between these two events, with there being more tremors than sounds. Any discussion of their correlation has to be effectuated through comparing their statistics. We thus grouped the nonzero values of the time derivative of force, $dF/dt$, into a histogram. A power-law distribution $P(dF/dt)\sim 1/(dF/dt)^{\beta}$ emerged with $\beta=1.31$  in Fig.\ref{FSR}.
This is to be contrasted to Fig.\ref{data}(a). The fact that both figures exhibit power-law behavior with similar exponent and valid range of parameter  confirms a positive correlation between energies of the quake and the crackling sound.
The chaotic distribution when $dF/dt$ exceeds 10N/s was caused by the drop of sensitivity in FSR when the force change is larger than 10N. Limited by the sensitivity of FSR, we believe the number of small events it picked up was underestimated which probably explains why $\beta=1.31$ is slightly smaller than $b=1.45$ in Fig.\ref{data}(a).

A minimal model will be presented to explain qualitatively why the major seismic laws should appear in the humble bamboo chopstick. In contrast  to numerical simulations that are popular for the study of fracture in disordered media\cite{minozzi}, we believe analytical derivations can provide more insights. Since the Gutenberg-Richter and B{\aa}th laws concern only the energetics, they will be tackled separately from the Omori's which requires dynamics. For simplicity, we only consider a rectangular cross section, but other shapes can be easily generalized. 

\begin{figure}[!h]
  \centering
  \includegraphics[width=8.5cm]{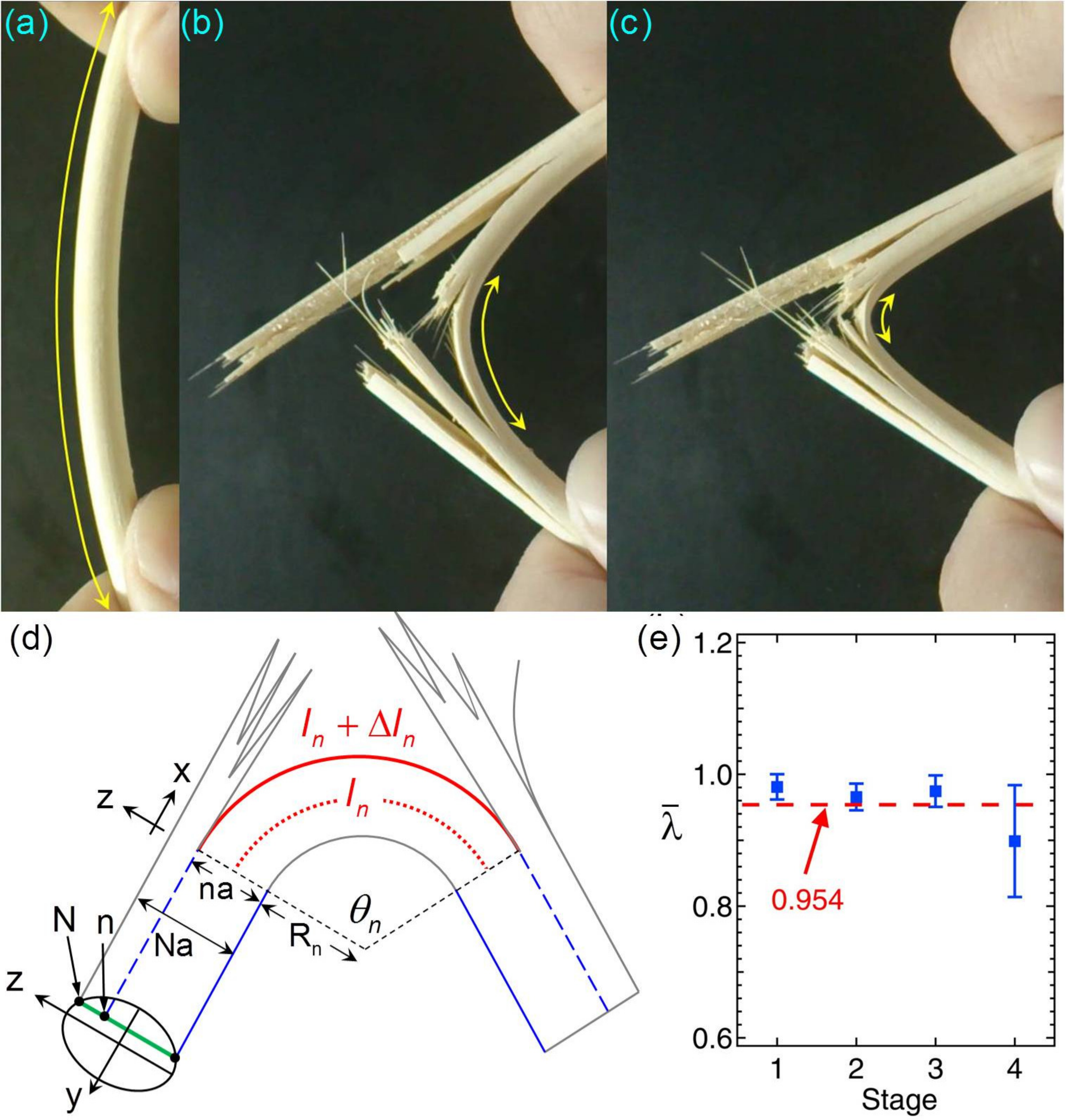} 
  \caption
  {(color online)  Yellow double-arrowed lines in panels (a,b,c) highlight the shortening of effective length. In panel (d), fiber on the $n$-th layer is stretched and on the verge of fracturing. The $R_n$ represents its radius of curvature, and $\theta_n$ the angle spanned by the bent section (in dotted red line). Note the end segments (in blue dashed line) remain straight, resembling the V-shape configuration in Ref.\cite{qiwa}. In panel (e), the  reduction ratio of effective length is determined for different time stages.
}\label{theory} 
\end{figure}

Two features in Fig.\ref{theory} merit attention. First is the progressive shortening of effective length, $l_n$ - the length of beam neutral axis that is actually bent. Second, the end segments of the chopstick remain straight because they are thicker and less susceptible to bending.  To nail down the relevant factors and enable analytic solutions, we will focus on the bending energy in the static beam equation, which includes tensional and compressional stresses in the cross section external and internal to the beam neutral axis, respectively:  
\begin{equation}
E_{\rm tot}\approx\int^{l_n}_0 dx\int dy  \frac{K_{B}}{2} \left(\frac{1}{R_{n}}\right)^{2}
\label{energy}
\end{equation}
where $K_{B}\sim Y(na)^{3}$ and
$Y$ denotes the Young's modulus. As exemplified by Fig.\ref{theory}(b, c), the fractured segment of bamboo remains rather straight, which demonstrates that the plastic region preceding fracture is very narrow and the elastic form of Eq.(\ref{energy}) can be extended to large strains. As fracture proceeds in the tensional region, $n$ decreases and $l_n$ is shortened progressively as evidenced by Fig.\ref{theory}(a, b, c) and
our YouTube clip\cite{youtube}. 

Geometric relations can be written down from Fig.\ref{theory}(d)
\begin{equation}
\begin{split}
& l_{n}+\Delta l_{n}=\left( R_{n}+na \right)\theta_{n} \\
& l_{n}=\left(R_{n}+\frac{na}{2}\right) \theta_{n}
\end{split}
\end{equation}
Subtracting them gives
\begin{equation}
\frac{na}{2}\cdot\theta_{n}=\Delta l_{n}=\varepsilon l_{n}
\end{equation}
where the threshold strain for fracture, $\varepsilon\equiv \Delta l_{n}/l_n$, is a material property.
Simple algebra gives
\begin{equation}
\frac{\Delta l_{n}}{l_{n}}=\frac{\frac{na}{2}}{R_{n}+\frac{na}{2}}\approx\frac{na}{2R_{n}}.
\label{thet}
\end{equation}
since $R_n\gg na$. For a rectangular cross section, the $y$-integration simply gives a constant and the energy stored in the $n$-th layer is roughly
\begin{equation}
E_{n}\approx E_{\rm tot}/n=\frac{1}{n}\cdot l_{n}\cdot\frac{Y\cdot na}{2}\cdot \varepsilon^{2}\propto l_{n}.
\label{stretching}
\end{equation}
Since the stress is larger away from the beam neutral axis, the equipartition approximation of $E_n/E_{\rm tot}\approx 1/n$ is an underestimation.

The fracture process was divided into four time stages in Fig.\ref{theory}(e). If there are 400 pulses, take the ratio of $l_{301}/l_{400}$, where $l_{301}$ and $l_{400}$ represent the effective length measured upon the 100th and the first pulses, and define it as ${\bar\lambda}^{100}$ where $\bar\lambda$ is the mean value. Figure \ref{theory}(e) shows that $\bar\lambda$ is roughly constant. The deviation of $\bar\lambda$ value and a large error bar in the final stage is likely caused by the uncertainty in  pulse number because crackling sound becomes very feeble. From these observations, the effective length can be assumed to obey
\begin{equation}
l_n=l\cdot\lambda^{N-n}
\label{ln}
\end{equation}
where $l$ denotes the initial length and $\lambda <1$ is a material property. Similar concept has been used to explain the power-law behavior of the crackling sound from a crumpled thin sheet\cite{tsai}.

The total number of crackling sound equals
\begin{equation}
\sum_n \approx \int dn=\int\frac{dn}{dE_{n}}dE_n.
\end{equation}
Divide both sides by $\sum$, we get unity on the left side to conserve probability while the occurrence rate can be identified as $P(E_n)\propto dn/dE_{n}\sim 1/E_n$ by use of Eqs.(\ref{stretching}, \ref{ln}). This reproduces the power law in Fig.\ref{data}(a). If we increase the ratio $E_n/E_{\rm tot}=1/n$, the exponent will become larger and closer to the empirical value $b=1.45$ in Fig.\ref{data}(a).  When the rectangular cross section is changed to, say, southern hemisphere of a horizontally halved chopstick, the width and fiber number of each layer decrease as fracture propagates downward from the top. Since top layers emit stronger pulses according to Eqs.(\ref{stretching}, \ref{ln}), there is now more count of loud sound than weak sound. And we expect Fig.3(a) to level off and give a smaller $b$. This is consistent with our observation. 

To derive the Omori's law, we need the time interval between successive aftershocks to calculate the elapsed time in Fig.3(b). By assuming a constant angular acceleration during this short interval, we can estimate this time, $\Delta t_n$ for the torque to render a displacement of $\Delta \theta_n$: 
\begin{equation}
\Delta\theta_{n}=\theta_{n-1}-\theta_{n}=\frac{1}{2}(Fl\sin\theta_{n}/I)(\Delta t_{n})^{2}
\label{angle}
\end{equation}
where $I$ denotes the moment of inertia. Since the bending force, $F$, comes from far ends of chopstick, the arm of force equals $l$, not $l_n$. The $\sin\theta_{n}$ factor is due to the fact that the force is not perpendicular to its arm. Finally, since $\theta_{n}$ is mostly small, $\sin\theta_{n}\approx \theta_{n}$. As $n\gg 1$, we can replace $\Delta\theta_{n}$ by $d\theta_{n}/dn$ and obtain from Eq.(\ref{angle}):
\begin{equation}
(\Delta t_{n})^2\propto \ln\lambda+\frac{1}{n}
\end{equation}
by use of Eqs.(\ref{thet}, \ref{ln}). This shows that the time interval to the next aftershock  increases with the number of broken fibers, $N-n$. A simple picture to understand this slowdown in braking is that the tensional stress external to the beam neutral axis is lessened, as the chopstick gets thinner. Consequently, it takes a larger angular displacement and longer time  to reach the threshold strain for fracture. The aftershock rate can be calculated by dividing the number of aftershocks, $n_{ms}-n$, by the time lapse from their corresponding main shock, $\Delta t_{n_{ms}-1}+\Delta t_{n_{ms}-2}+\cdots+\Delta t_{n}$ where $n_{ms}\gg 1$ labels any one of the main shocks. Since the maximum number of aftershocks, $n_{ms}-n$, is in the ballpark of 8, we can Taylor-expand the denominator by  the small parameter, $1-(n/n_{ms})\ll 1$ and obtain
\begin{equation}
{\rm rate}=\frac{n_{ms}-n}{\sum^{n_{ms}-1}_{n}\Delta t_{n}}\approx \frac{n_{ms}}{D'+(D''/2)[1-(n/n_{ms})]}
\label{omori}
\end{equation}
 where the $D'$ and $D''$ denotes the first and second derivative. Since $n_{ms}-n$ is roughly a measure of time, Eq.(\ref{omori}) reproduces the  shift and power law of Fig.\ref{data}(b).

Since $E_n/E_{n-1}\sim 1/\lambda$ is independent of the value of $n$ and insensitive to the approximation of $E_n/E_{\rm tot}=1/n$, the B{\aa}th law is naturally guaranteed. 

In conclusion, we establish a complete parallel between the crackling noise of a common bamboo chopstick and a bundle of spaghetti and the fundamental seismic laws. The statistics of these acoustic events is shown to correlate with that of tremor. The fitting function and corresponding parameters are determined by the rigorous statistical method of Akaike Information Criterion\cite{tsai,aic} and MLE. We succeed at deriving these  laws analytically without invoking the concept of phase transition\cite{sethna}, self-organized criticality\cite{soc,bak}, or fractal\cite{fractal}. Our models are based mainly on a structured cross section, which can be either fibrous or layered. This shift of emphasis from mechanics to geometry to explain the power-law behavior is in conformance with the proposal of Ref.\cite{carpinteri}. 

We gratefully acknowledge funding from MoST in Taiwan, technical supports by Jing-Ren Tsai, and the hospitality of the Physics Division of NCTS in  Hsinchu.

\end{document}